\documentclass{article}

\usepackage{graphicx}
\usepackage{psfig}
\usepackage{epsfig}
\usepackage[round]{natbib}

\title{The Assessment and Calibration of Ensemble Seasonal Forecasts of Equatorial Pacific Ocean
Temperature and the Predictability of Uncertainty}

\author{Stephen Jewson$^1$\footnote{\emph{Correspondence address}: RMS, 10 Eastcheap,
London, EC3M 1AJ, UK. Email: \texttt{x@stephenjewson.com}}, Francisco Doblas-Reyes$^2$ and Renate Hagedorn$^2$\\
$^1$RMS, London, United Kingdom\\
$^2$ECMWF, Reading, United Kingdom}

\begin{document}

\maketitle

\begin{abstract}

We evaluate the performance of two 44 year ensemble seasonal hindcast time series 
for the Ni\~{n}o3 index produced as part of the DEMETER project. 
We show that the ensemble mean carries useful information out to six months.
The ensemble spread, however, only carries useful information out to four months in one of the
models, and two months in the other.

\end{abstract}

\section{Introduction}

One of the ways that users of seasonal forecasts can access forecast information is to monitor predictions of sea
surface temperatures in the equatorial Pacific. The Ni\~{n}o3 and Ni\~{n}o3.4 regions in particular are good indicators
of the state of ENSO, and numerous forecasts are available for these indices. These forecasts come 
from dynamical ensemble
models such as those discussed in~\citet{stockdaleaab} and~\citet{masonggysa}, 
from intermediate complexity models as described in~\citet{prigaudcdfn} and from pure statistical 
models such as that of~\cite{penlandm}.

Users of any of these forecasts need to be able to answer a number of questions about a forecast before they can use
it with confidence in an application. These questions include:

\begin{itemize}
  \item Does the forecast need calibration?
  \item For how many months is the forecast of the mean temperature better than no forecast?
  \item What is the best way to derive a prediction of the uncertainty around the mean?
  \item Which is the best of the available forecasts?
\end{itemize}

These questions have been addressed by a number of authors.
Methods used to answer the first question include assessments
of bias and incorrect variance~\citep{doblasreyesps} and rank histograms and reliability diagrams~\citep{wilks00}. 
The second question
can be answered using the anomaly correlation between forecast and observations or the significance
level of a regression coefficient between forecast and observations~\citep{brankovicp}. 
The third question is generally harder
to answer and there do not seem to be any very satisfactory methods described in the literature.
Finally, the fourth question can be answered using a combination of methods including anomaly correlation or
mean squared error (MSE), which both address the skill of the mean of the forecast, 
or the relative operating characteristic (ROC)~\citep{swets88} and CRPS (continuous rank probability
score)~\citep{hersbach} which look at aspects of the distribution.

As part of a project to develop simple practical methods for use in industry,
we will now show how it is possible to answer all four of these questions at once in a single
consistent framework using the method described in~\cite{jewsonbz03a} (henceforth JBZ).
The values of the parameters from the JBZ method will give us clear answers to the first three questions, 
along with interesting insights into the predictability of uncertainty. We will answer the fourth
question using a new skill measure that measures the ability of the forecast to predict
the whole temperature distribution. 

In section~\ref{data} we describe the observational and forecast data used for this study. 
In section~\ref{method} we will briefly
describe the method we will apply and the interpretation of the resulting parameter values.
In section~\ref{results} we show results from the analysis of our two time series of ensemble
seasonal hindcasts, and in section~\ref{conclusions} we discuss our results and draw some conclusions.

\section{Data}
\label{data}

All the analyses in this paper are based on sets of seasonal hindcasts from the European Centre for Medium
Range Weather Forecasting (ECMWF) and Meteo France (MF) seasonal forecast models. 
These are both dynamical ocean-atmosphere models. The hindcasts were produced as part of the DEMETER
project~\citep{demeter} and consist of 6 month predictions of monthly mean Ni\~{n}o3 temperatures. 
Both sets of hindcasts cover the period 1958 to 2001 (44 years) and consist of ensembles of size nine. 
The forecasts start at four different times of year: February, May, August and November. 
These forecasts are compared with temperature observations for the Ni\~{n}o3 region obtained from the
GISST2.3 and Reynolds 2D-Var SST data sets.

All values used in this study are deseasonalised in the mean and the standard deviation before any 
calculations are performed, and all equations given below apply to these deseasonalised values.

\section{Method}
\label{method}

We will address the four questions outlined in the introduction using the statistical 
model described in JBZ.
This model consists of a simple calibration method in which the ensemble mean and the ensemble
standard deviation are considered as inputs for a prediction for the mean temperature and the 
uncertainty around that temperature. The predictions of the mean and the uncertainty are derived from the ensemble
mean and standard deviation using simple linear transformations. 
To be more specific, the model postulates that observed temperatures come from a normal distribution with 
estimated mean given by $\hat{\mu}_i=\alpha+\beta m_i$ and estimated standard deviation given by 
$\hat{\sigma}_i=\gamma+\delta s_i$, where $m_i$ is the ensemble
mean and $s_i$ is the ensemble spread. We write this model as:

\begin{equation}\label{LEC}
 T_i \sim N(\alpha+\beta m_i, \gamma+\delta s_i) = N(\hat{\mu},\hat{\sigma})
\end{equation}

Although neither the mean temperature nor the standard deviation of temperature are actually observed, 
the parameters of this model can be fitted easily by maximizing a cost function which measures the goodness
of fit of the normal distribution. There are a number of possible cost functions one could use, but
we choose to use the likelihood, defined as the probability density of the observations given the calibrated forecast.
Likelihood is the standard approach used for parameter estimation in statistics 
(see for example the statistics textbooks by~\citet{casellab02} or~\cite{lehmannc98})
and is one of the most natural methods for measuring the goodness of fit of a distribution.
It also gives the most accurate possible parameter estimates for most statistical models.

Having determined the optimum values for the parameters $\alpha, \beta, \gamma$ and $\delta$ one can
interpret them as follows:

\begin{itemize}

  \item $\alpha$ identifies bias in the forecast.
  If $\alpha$ is significantly different from zero then the bias needs correcting.

  \item $\beta$ calibrates the variations of the ensemble mean to have the correct amplitude. 
  If $\beta$ is significantly different from one then the forecast needs calibration.
  It is also used to assess whether the forecast contains any useful information: if $\beta$ is not significantly different
  from zero then the forecast is useless (at least in the context of this calibration model) 
  and one should use climatological temperatures instead.

  \item $\gamma$ and $\delta$ calibrate the prediction of the uncertainty to have the correct
  size and variability.
  If there is very little information in the variability of the ensemble spread, 
  then $\delta$ will be small and $\gamma$
  will be larger to compensate. In the same way that $\beta$ can be used to assess whether the ensemble
  mean has any useful predictive information, $\delta$ can be used to assess the ensemble spread. 
  If $\delta$ is not significantly different from zero then the ensemble spread contains
  no useful information (again, in the context of this model).

\end{itemize}

In the case in which $\delta$ is not significantly different from zero then, 
for that lead time, the hope that flow-dependent variations in uncertainty can be predicted
has to be abandoned and one should re-fit using the alternative model: 
  
\begin{equation}\label{reg}
 T_i \sim N(\alpha+\beta m_i, \hat{\sigma}_0)
\end{equation}
 
where $\hat{\sigma}_0$ is constant in time for a given lead time.
This model can also be fitted by maximizing the likelihood, which is equivalent to 
least squares linear regression in this case.
Predictions of the uncertainty in this simplified model will come entirely from past forecast error statistics. 

Even if $\delta$ is significantly different from zero and we conclude that the ensemble spread of a 
model does contain useful flow-dependent information about variations in 
the uncertainty, it is still not necessarily the
case that the predicted variations in the uncertainty are material. In other words, they may be so small 
relative to the mean level of the uncertainty that they can as well be ignored. This can be for two reasons:
either because the predicted variations in the uncertainty have a very low correlation with the actual 
variations in the uncertainty, or because the actual uncertainty itself simply does not vary very much.
We will measure the materiality of the predictable variations in the uncertainty using the coefficient of variation 
of the spread (COVS), defined by the ratio of the standard deviation of the variations in the 
calibrated spread to the mean calibrated spread:

\begin{equation}
  \mbox{COVS}= \frac {\delta \mbox{sd}(s)} {\gamma + \delta \overline{s}}
\end{equation}

We arbitrarily choose a level of 0.05 to define the level below which we consider variations in the uncertainty to
be immaterial. If uncertainty variations are immaterial then there is no need to use equation~\ref{LEC} to calibrate a
forecast, and the simpler linear regression model (equation~\ref{reg}) can be used instead.

There are a number of ways to compare the resolution of forecasts from different models. To make a fair
in-sample comparison, the same calibration methods must be used for all the models. Forecasts can then 
be compared using methods such as MSE, ROCs, or CRPS. However, to focus on the ability of a model to
predict the probabilities across the whole distribution correctly we prefer to use the likelihood.
Likelihood skill measures can be presented in a number of ways:

\begin{itemize}
  \item The likelihood itself. 
  \item The log-likelihood, which has a more compressed (and hence more convenient) range of values
  than the likelihood
  \item The square root of minus the log-likelihood (RMLL). This has the advantage that it is equivalent to use
  of the root-mean-square error in cases where the uncertainty is not flow-dependent.
  \item The log-likelihood skill-score (LLSS) defined as one minus the ratio of the log-likelihood to the
  climatological log-likelihood. This has the advantage that values range from zero 
  for a useless forecast to one for a perfect forecast.
\end{itemize}

Likelihood skill measures can be presented for each lead time, or for all lead times together. 
We will present our comparison results in terms of the LLSS for each lead time.

\section{Results}
\label{results}

\begin{figure}[!htb]
  \begin{center}
    \scalebox{0.9}{\includegraphics{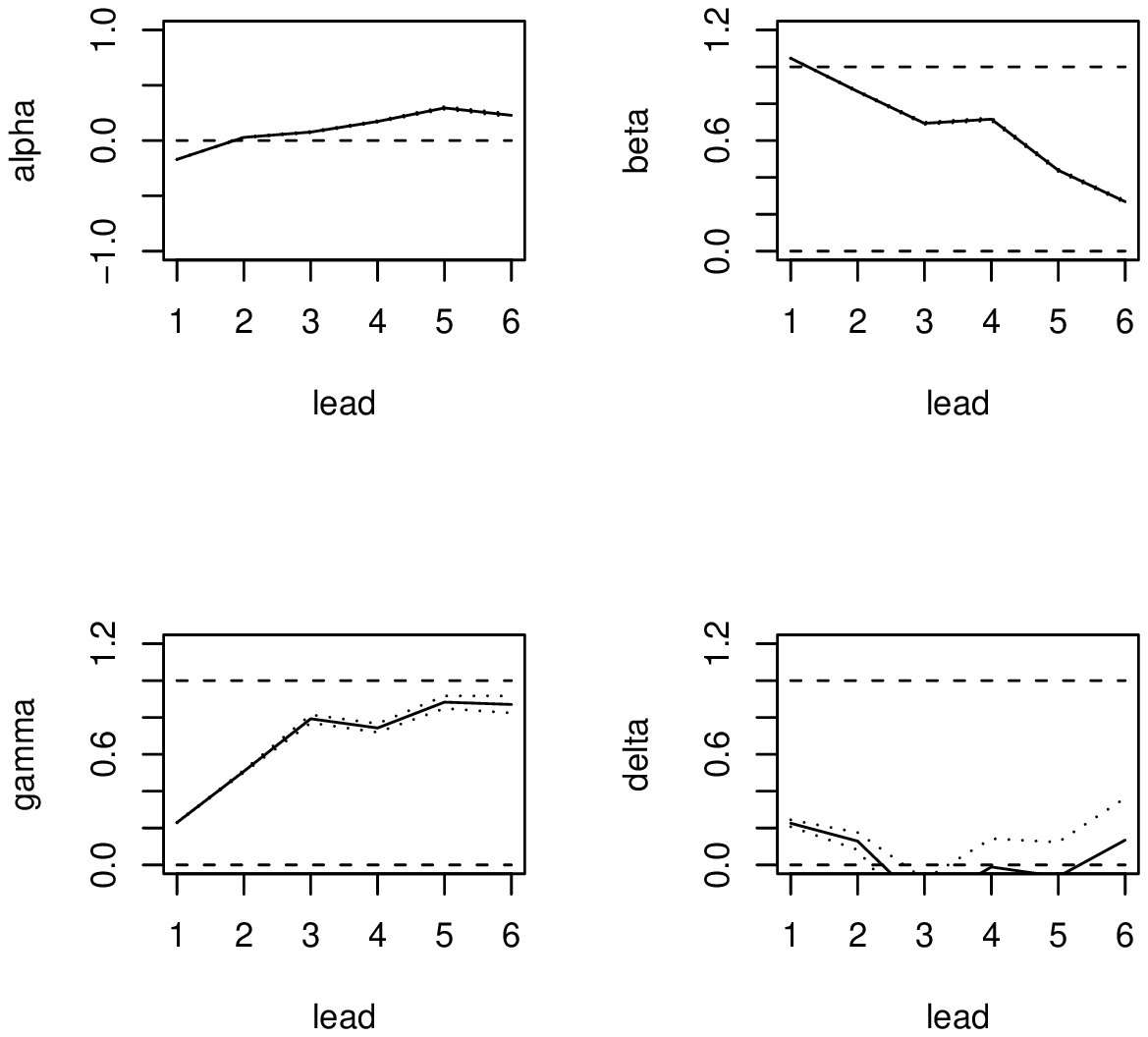}}
  \end{center}
 \caption{The optimum values of the parameters in equation~\ref{LEC} for the ECMWF seasonal hindcasts,
 calculated using forecasts made at all times of year.} 
 \label{ecmwfannual}
\end{figure}

Figure~\ref{ecmwfannual} shows the optimum values for the parameters in equation~\ref{LEC} for the ECMWF hindcasts,
based on forecasts made at all times of year. Each estimate also has an indication of the statistical 
or sampling uncertainty around the optimal parameter estimate. These uncertainties are calculated from the
curvature of the log-likelihood in the standard way.
44 years of forecasts four times per year gives 176 past forecasts, and from the uncertainty estimates we see that this
is enough to give reasonably accurate estimates of all the parameters. 

The value of $\beta$ is significantly
different from zero at all lead times up to six months. This shows that the ensemble mean 
contains useful information at all lead times. 
However, it certainly needs to be calibrated to give an optimum forecast, especially at the longer lead times.
The value of $\delta$, however, is only significantly different from zero at leads one and two. This shows that the 
ensemble spread only contains useful information at these two lead times, and for longer leads does not contain
useful information. For leads three to six one thus has to discard the results from the calibration model and 
refit the ensemble
data using standard linear regression (equation~\ref{reg}). 
This will give optimal predictions of the mean and the uncertainty, but
the uncertainty prediction will not be flow-dependent. 

\begin{figure}[!htb]
  \begin{center}
    \scalebox{0.9}{\includegraphics{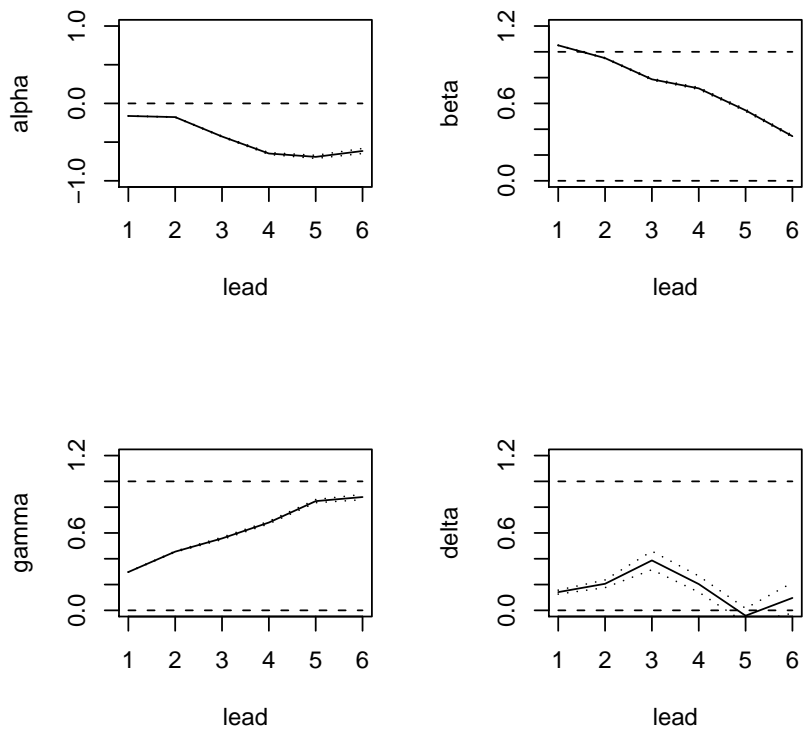}}
  \end{center}
 \caption{As for figure~\ref{ecmwfannual} but for the MF seasonal hindcasts.} 
 \label{mfannual}
\end{figure}

Figure~\ref{mfannual} shows results for the same analysis for the MF model. The value of $\beta$ again shows that
the ensemble mean contains useful information at all lead times. The values for $\delta$ are now significantly 
different from zero up to lead four. Only at leads five and six does the ensemble spread contain no useful information.
Once again, at these lead times the results from the calibration model must be discarded and a linear regression model 
used instead.

We have seen that the ECMWF model can be used to make a flow dependent prediction of uncertainty at leads one and two,
and the MF model can be used to make such a prediction for leads one to four. We now assess the size of these
flow-dependent variations in uncertainty relative to the mean level of the uncertainty using the COVS. 
Values for the COVS for leads for which there
is significant information about the spread are given in table~\ref{covs}. We see that the ECMWF model only
gives material values of COVS at lead one. However at this lead the standard deviation of predictable variations 
in the uncertainty is nearly 14\% of the total uncertainty. 
It would seem likely that this is a useful level of predictability of the variability of uncertainty for some users.
The MF model gives material but rather low values for the COVS out to lead four.

We now address the question of whether the models show the same levels of predictability 
at different times of year by repeating the calibration analysis on each season separately. 
The seasonal parameters are fitted using only 44 past forecast values (one per year) and this means that
the parameters cannot be estimated as accurately. This is balanced by the fact that we can
hopefully pick up more detail in the structure of the predictability. 
For instance we might expect to see some signs of the well-known
spring barrier~\citep{webstery}.

\begin{figure}[!htb]
  \begin{center}
    \scalebox{0.9}{\includegraphics{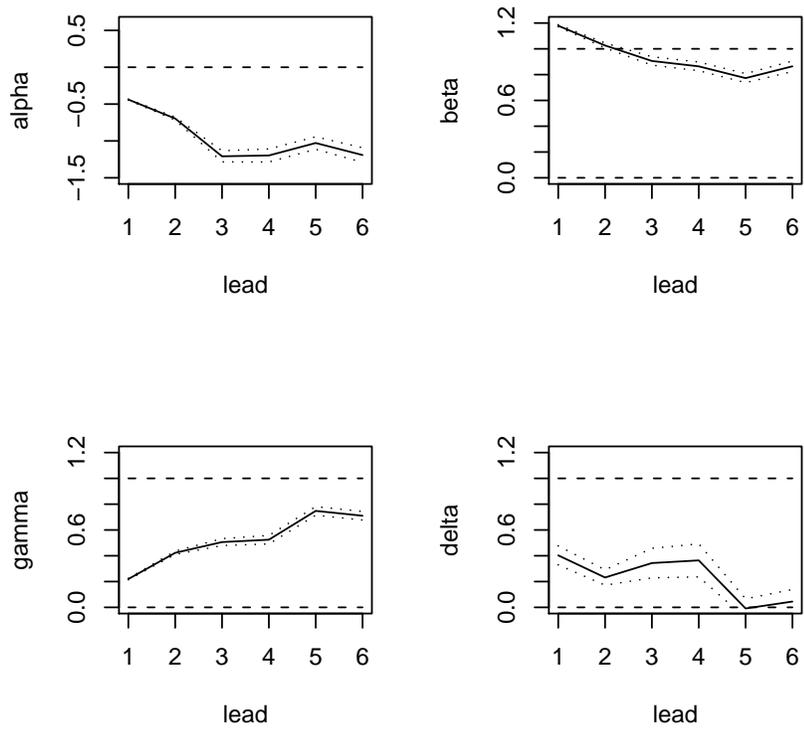}}
  \end{center}
 \caption{As for figures~\ref{ecmwfannual} and~\ref{mfannual} but now based on MF hindcasts started in May only.} 
 \label{mfforecast1}
\end{figure}

The results for the ensemble mean (not shown) show that in all seasons both models contain
useful predictive information out to six months. The results for the ensemble spread are more complex.
Figure~\ref{mfforecast1} shows the optimum values for the parameters in equation~\ref{LEC}
for the MF forecasts started in May.
In this case the ensemble spread contains detectable useful information from leads one to four, as with the
annual data. 
Results for the uncertainty for all other seasons and for both models are summarized in table~\ref{seasonal}, which
lists the months for which skillful flow dependent predictions of the variations in uncertainty can be made. 
We detect that the ECMWF model shows skill for the spread only in the first month, while the MF model
shows skill for the spread for different numbers of months in each season. 
How is it that, when analyzed on annual data, the ECMWF model shows two months of significant spread,
whereas when analyzed on seasonal data, it only shows one month? The answer is that the spread signal in the 
second month is very small, and reducing the number of data points from 176 to 44 means that it can no longer
by detected. 
The MF model shows more seasonal variability in the predictability of spread than the ECMWF model.
In particular, the 
model shows the least predictability of spread in February. This suggests that the spring predictability
barrier may affect predictions of spread as well as predictions of the mean. 

\begin{figure}[!htb]
  \begin{center}
    \scalebox{0.9}{\includegraphics{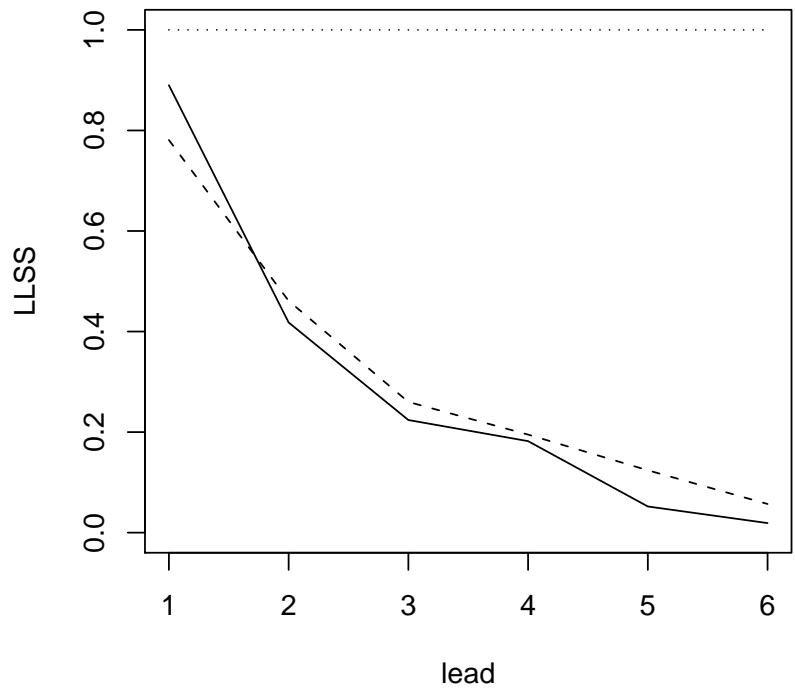}}
  \end{center}
 \caption{The log-likelihood skill score (LLSS) for the ECMWF seasonal hindcasts (solid line) and the
 MF seasonal hindcasts (dashed line). A zero score indicates no more information than climatology,
 and a score of one indicates a perfect forecast.} 
 \label{compare}
\end{figure}

Finally we consider which of the two models produces better forecasts, taking into account
the skill with which the model predicts the probabilities across the whole distribution by using a 
likelihood based skill measure.
Figure~\ref{compare} shows the LLSS for both models.
We see that, by this measure, the ECMWF model is better at lead 1, but that the MF model is slightly better at all
subsequent leads. The biggest fractional differences in the LLSS are at leads five and six. 
However, the differences are small and may be partly due to sampling variability.

\section{Conclusions}
\label{conclusions}

We have described how the ensemble calibration model of JBZ 
can be used to assess and calibrate
ensemble seasonal forecasts, and in particular how it can determine the limit of useful information in the
ensemble mean and standard deviation of such forecasts.
We have assessed time series of ensemble seasonal hindcasts from the ECMWF and MF models. 
The results are striking.
For the ECMWF model although the ensemble mean contains useful predictive information out to the end of
the forecast period at six months, the ensemble spread shows no useful predictive information beyond
lead two. At lead times
beyond the second (and beyond the first on a seasonal basis) 
it is more appropriate to calculate the uncertainty of the forecast from past forecast error
statistics than it is to use the ensemble spread. Even at lead two, the size
of the predictable component of the uncertainty could be considered immaterial relative to the total uncertainty.
This finding contrasts with the results
for \emph{medium range} ensemble forecasts from ECMWF, for which both the mean and the spread contain useful 
information out to the end of the forecast at 10 days (see JBZ).
For the MF model the ensemble mean also contains useful information to the end of the forecast, while the 
spread contains useful information up to month four. Only at months five and six does the spread not contain useful
information about the variations in forecast uncertainty. When we consider the size of the variations in spread
that are predictable from the MF model, we find that they are material relative to the total uncertainty, and 
are probably worth incorporating into a prediction of the uncertainty if accurate predictions of the uncertainty
are important. They are, however, only a small fraction of the mean uncertainty and some users may decide to ignore
them if less accuracy in the estimates of the uncertainty is required.
When analyzed on a seasonal basis we see strong seasonal variability in the predictability of
uncertainty consistent with there being a spring barrier for predictability of uncertainty as well as for
predicting the mean.

We have demonstrated how long seasonal hindcasts are extremely useful both for defining accurate calibration parameters
and for assessing the skill of forecast systems. A prudent user of forecasts would never use a forecast unless
the skill can be identified statistically in hindcasts or past forecasts, 
and with seasonal forecasts skill can often only be
identified with such long data sets. 

Finally we have compared the ECMWF and MF hindcasts using a likelihood based skill measure which assesses 
the performance
of the model in predicting the whole distribution of possible outcomes. We find that, according to this measure, the
ECMWF model makes slightly better forecasts at lead one, while the MF model forecasts are slightly better 
for leads two to six.

There are a number of directions for future work arising from this study.
It is important to assess results from calibration using the JBZ  model
in out-of-sample tests, and to 
compare the reliability of such results with those from other calibration methods. From a user-perspective
it is important to extend this work to consider site-specific forecasts in addition to forecasts of the
Pacific ocean temperature. Finally, the optimal predictions of the mean and variance that are produced by the
calibration method described form a good basis for forming optimal multimodel forecasts, 
and it would interesting to compare the results
from such multimodel forecasts with those derived using other methods.

\section{Acknowledgements}

We would like to thank ECMWF for providing us with the data on which this study was based,
and D. Anderson, A. Brix, S. Mason and C. Ziehmann for helpful discussions.
SJ funded his own research, while FDR and RH were funded by the DEMETER project (EVK2-1999-00197).

\bibliography{jewson}

\clearpage

\clearpage

\clearpage
\begin{table}[!htb]
\begin{center}
\begin{tabular}{|c|c|c|c|c|}
 \hline
 \hline
lead        &    1     &   2     &   3   &    4      \\ \hline
ECMWF       &   13.9   &   4.9   &   N/A &    N/A    \\ \hline
MF          &   8.6    &   8.8   &  12.0 &    5.3    \\ \hline
\end{tabular}
\end{center}
\caption{The values, as percentages, for the coefficient of variation of the predictable spread from the
ECMWF and MF models. Values for the ECMWF hindcasts are only given up to lead two since beyond that
there is no detectable skill in predicting the spread. Similarly, values for the MF hindcasts
are given up to lead four only.
}
\label{covs}
\end{table}

\clearpage

\clearpage
\begin{table}[!htb]
\begin{center}
\begin{tabular}{|c|c|c|c|c|}
 \hline
 \hline
forecast month                & Feb & May & Aug &  Nov     \\ \hline
ECMWF          &   1   &   1   &   1   &    1      \\ \hline
MF             &   0   &   4   &   1   &    2      \\ \hline
\end{tabular}
\end{center}
\caption{
The number of months for which the forecast started in \emph{forecast month} 
showed useful predictive information in the ensemble spread, as judged by
a value for the parameter $\delta$ in equation 1 being significantly different from zero.
}
\label{seasonal}
\end{table}

\clearpage

\end{document}